\documentclass[prb, preprint, letter, superscriptaddress]{revtex4}
\usepackage{amssymb}
\usepackage{amsmath}
\usepackage{bbm}
\usepackage{graphicx}
\usepackage{color}

\newcommand{\ris}{\{\vec r_i\}}
\newcommand{\rips}{\{\vec{r'}_i\}}
\newcommand{\Lne}{\hat L_{_\textrm{NE}}}

\newcommand{\rip}{\vec{r'}_i}
\newcommand{\rjp}{\vec{r'}_j}
\newcommand{\ri}{\vec{r}_i}
\newcommand{\rj}{\vec{r}_j}
\newcommand{\hrij}{\hat r_{ij}}

\begin{document}

\author{Shenshen Wang}
\author{Peter G. Wolynes}
\affiliation{Department of Physics, Department of Chemistry and Biochemistry,  and Center for Theoretical Biological Physics, University of California, San Diego, La Jolla, CA 92093, USA}
\affiliation{Department of Chemistry, and Center for Theoretical Biological Physics, Rice University, Houston, TX 77005, USA}

\title{Active contractility in actomyosin networks}
\date{\today}

\begin{abstract}

Contractile forces are essential for many developmental processes involving cell shape change and tissue deformation. Recent experiments on reconstituted actomyosin networks, the major component of the contractile machinery, have shown that active contractility occurs above a threshold motor concentration and within a window of crosslink concentration. We present a microscopic dynamic model that incorporates two essential aspects of actomyosin self-organization: the asymmetric load response of individual actin filaments and the correlated motor-driven events mimicking myosin-induced filament sliding. Using computer simulations we examine how the concentration and susceptibility of motors contribute to their collective behavior and interplay with the network connectivity to regulate macroscopic contractility. Our model is shown to capture the formation and dynamics of contractile structures and agree with the observed dependence of active contractility on microscopic parameters including the contractility onset. Cooperative action of load-resisting motors in a force-percolating structure integrates local contraction/buckling events into a global contractile state via an active coarsening process, in contrast to the flow transition driven by uncorrelated kicks of susceptible motors.

\end{abstract}

\hyphenation{}

\maketitle

\section{Introduction}


Contractile forces are essential for many processes vital to development, ranging from cytokinesis and cell motility \cite{cell motility} to wound healing and gastrulation \cite{gastrulation}. Networks of filamentous actin (F-actin) and the molecular motor, type II myosin have been identified as the major components of the contractile machinery. The actin network provides a structural scaffold on which the myosin motors move, powered by ATP hydrolysis. Actomyosin networks generate contractile forces through the activity of myosin motors, which themselves assemble into bipolar minifilaments that generate sustained sliding of neighboring actin filaments relative to each other in order to reorganize F-actin networks and generate tension.\cite{morphogenesis} When coupled to the cell substrate or using cell-cell adhesions, contractile actomyosin networks transmit forces to their environment.

In addition to the microtubule-kinesin system, another important filament-motor assembly in cells that forms well-focused mitotic spindle poles driven by polarity sorting mechanism \cite{polarity sorting 1, polarity sorting 2} to accomplish high-accuracy segregation of duplicated chromosomes, actomyosin condensates appear in diverse tissues and organisms as transient structures that coalesce into still larger arrays that exert contractile forces. Examples include the contractile rings driving cytokinesis and wound healing, and the contractile networks that deform epithelial cell layers in developing embryos and drive polarizing cortical flows \cite{cortical flow 1, cortical flow 2}.

Some recent theoretical efforts have modeled the contractile actin cortex as an active polar gel and have derived effective continuum theories within a hydrodynamic framework. \cite{active polar gel 1, active polar gel 2, force dipole,KJ 03PRE, Lau 03PRL, Liverpool Marchetti 03PRL}
These macroscopic approaches based on generic symmetry considerations predict the formation of diverse patterns in acto-myosin gels such as asters and ring-like structures, which have been observed in studies \textit{in vitro}.\cite{active patterning}

Recently Bendix and coworkers have reconstituted contractility in a simplified system of F-actin, muscle myosin II motors, and $\alpha$-actinin crosslinks.\cite{active contractility} The well-controlled nature of this in vitro system allows a systematic study of the dependence of contractility on \emph{microscopic} parameters, such as the number and activity of myosin motors, crosslink density and actin network connectivity. It has been shown that contractility occurs above a threshold motor concentration and within a \emph{window} of crosslink concentrations.
Whereas earlier experiments on purified actomyosin solutions have established that contraction of F-actin networks by myosin II motors at physiological ATP level requires the presence of F-actin crosslinks, \cite{crosslinking 1, crosslinking 2} as has been confirmed by recent theoretical work,\cite{Carlsson 06PRE} this newly observed non-monotonic dependence of contraction tendency on crosslinking strength still calls for explanation.

There are two important aspects in actomyosin self-organization:
(1) Actin filaments have a highly asymmetric response to axial loading: They strongly resist tensile forces but easily buckle under compressive loads of several piconewtons. The ability to sustain large tension allows the motor-induced stresses to propagate significant distances through the network, whereas the buckling instability promotes formation of local actomyosin aggregates that coalesce into larger arrays exerting contractile forces (as exemplified in the spontaneous formation of myosin foci starting from a uniform distribution in an isotropic actin network \cite{Koenderink 09PNAS, multistage coarsening}).
(2) Motor-induced movements come in correlated pairs: the motor acts on a pair of parallel filaments to slide them past one another, inducing a pair of equal and oppositely directed moves at the crosslinks which are thus pulled together.
Despite the success of hydrodynamic theories in predicting diverse patterns,
a more microscopic model that could capture the nonlinear buckling behavior and the correlated motor-driven events, both of which are crucial for actomyosin self-organization and active contractility, is needed.


Here we provide a microscopic dynamic model for active contractility that combines the motor-driven stochastic processes (modeled as correlated kicks on motor-bonded crosslinks) with the asymmetric load response of individual actin filaments. This minimal model will be shown to exhibit the experimentally observed dependence of macroscopic contraction on motor concentration and actin network connectivity. The model also highlights the key role, in structural development, of motor susceptibility, a parameter characterizing how sensitively the motors respond to imposed forces.

\begin{figure}[htb]
\centerline{\includegraphics[angle=0, scale=0.52]{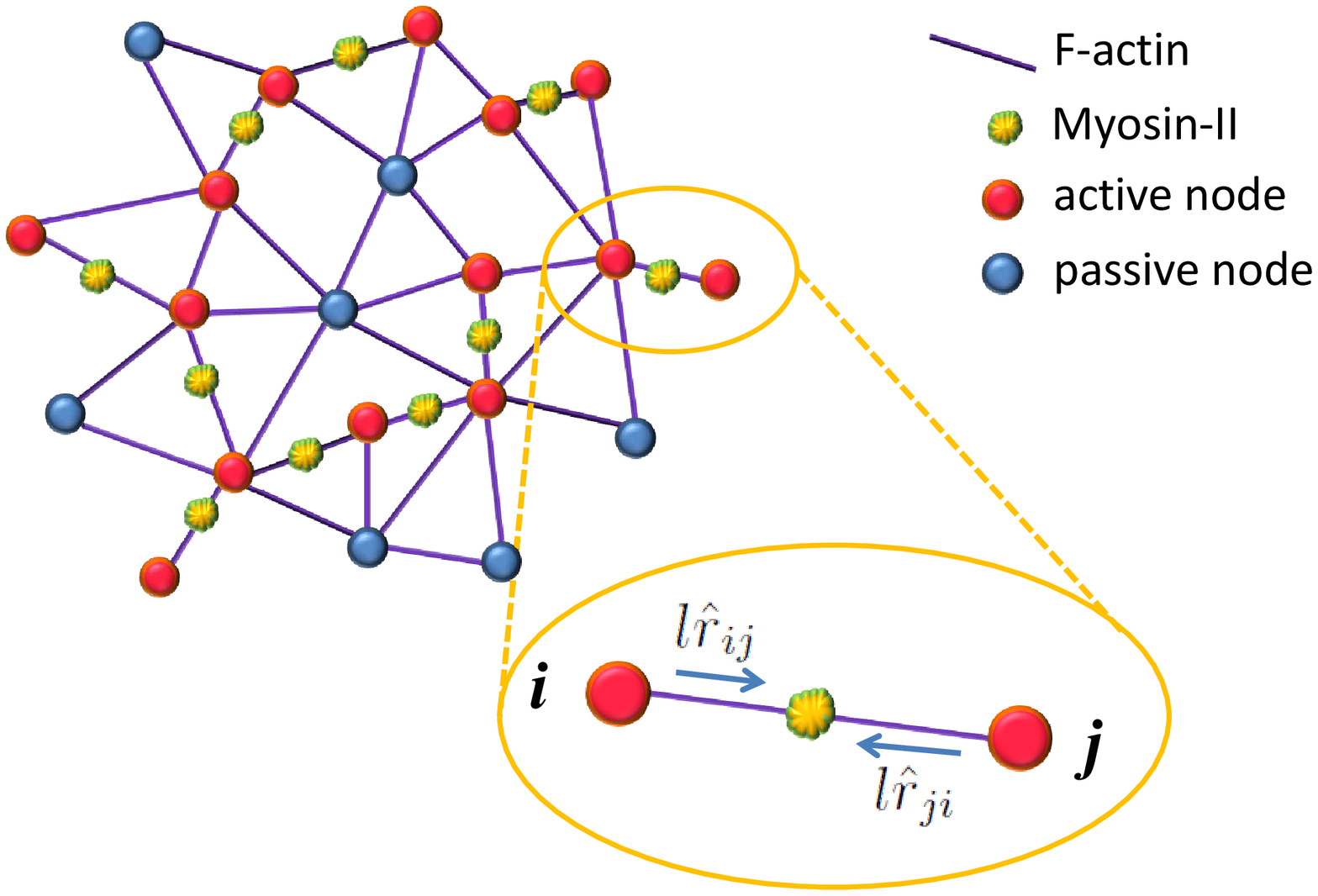}}
\caption{Schematic of the model system: A cat's cradle composed of a three dimensional amorphous network of crosslinked nonlinear-elastic bonds (purple lines) where motor-driven anti-correlated kicks induce pulsed local bond contraction and node aggregation. Spheres represent nodes/crosslinks and the yellow fuzzy objects stand for myosin II motor proteins. The size of the nodes and motors is exaggerated; excluded volume interaction is not implemented in current simulations. Red nodes are active nodes having motor-attached bonds. These are subject to motor kicks. Blue nodes are passive nodes only undergoing Brownian motion. An enlarged view of a unit of local contraction (circled) shows a pair of motor-bonded crosslinks undergoing anti-correlated kicks with fixed step size $l$ along the line connecting their centers where $\hrij$ is a unit vector pointing from node $i$ to node $j$.}
\label{schematic}
\end{figure}

By performing dynamic Monte Carlo simulations, we investigate the formation and dynamics of nonequilibrium structures in an actomyosin network modeled as a ``cat's cradle" \cite{CC, JCP} consisting of crosslinked nonlinear-elastic filaments subject to anti-correlated kicks on motor-bonded crosslinks (see Fig.~\ref{schematic} for a schematic illustration).
We first study how the concentration and susceptibility of motors determine the collective behavior generating diverse patterns.
We further construct a phase diagram for active contractility as a function of motor concentration and network connectivity at a given motor susceptibility. This diagram identifies the threshold motor concentrations and contains a window of network connectivity for active macroscopic contraction, consistent with observation.\cite{active contractility} We also find that at high connectivity contraction can still occur (at intermediate motor concentrations) but only if the excluded volume effect is negligibly small.
We finally compare the structures that develop for systems with correlated kicks to those when there are only uncorrelated kicks as in our earlier work \cite{spontaneous motion}. In particular, under uphill-prone motor kicks, the formation of ``asters" is replaced by the formation of disordered condensates resulting from an active multistage aggregation process due to buckling of connected actin structures induced by cooperative anti-correlated kicks. This prediction receives support from recent in vitro experimental studies on how the collective action of myosin motors organizes actin filaments into contractile structures.\cite{multistage coarsening}


Multiple factors that cause dynamic remodeling in cells are clearly absent in the reconstituted assay,\cite{active contractility} such as the disassembly of contractile structures and the transience of physiological actin cross-linking proteins.
In our numerical model, we likewise assume that network connectivity and motor distribution are quenched once initially assigned, in line with the fact that \textit{in vitro} structures are irreversibly assembled because many \textit{in vivo} factors allowing fast pattern renewals are left out. Therefore the nonequilibrium dynamics and structures exhibited in our numerical study arise solely from the intrinsic activity of motors firmly built into the actin network driving correlated movements stochastically. This contrasts with the two-fluid model that treats the cytoskeletal network as an elastic continuum where motor (un)binding kinetics leads to enhanced low-frequency stress fluctuations.\cite{force dipole}
The discreteness of the power strokes and thus the kick steps in our model bears relevance to the pulsed contraction observed in actomyosin networks \textit{in vivo}.\cite{contractile ratchet}


\newpage

\section{Model}

\subsection{Model system and dynamic rule}

We model the actomyosin network as a cat's cradle \cite{CC, JCP} consisting of nonlinear elastic filaments built on a three dimensional random lattice of volumeless crosslinks (Fig.~\ref{schematic}). These nonlinear filaments stretch elastically with effective stiffness $\beta\gamma$ when their contour length $r$ exceeds the relaxed length $L_e$ but buckle and become floppy upon shortening, as described by the pair interaction potential between bonded neighboring crosslinks $\beta U(r)=\Theta(r-L_e)\beta\gamma(r-L_e)^2/2$ where $\beta=1/k_BT$ and $\Theta(\cdot)$ is the Heaviside step function. The assumed weakness of the excluded volume effect allows large-scale structural rearrangements. We assign two mean field parameters to characterize the architecture of this filament-motor assembly: (1) the network connectivity, $P_c$, which denotes the fraction of nearest-neighbor pairs of crosslinks bonded by filaments; (2) the motor concentration, $P_a$, which indicates the fraction of active bonds, i.e. those attached by motors which induce equal and oppositely directed kicks on the connected crosslink pair. Network connectivity and motor concentration determine the fraction of active nodes against passive nodes: active nodes have motor-attached bonds and are subject to anti-correlated kicks with their motor-bonded neighbors whereas passive nodes have no motor-attached bonds and only undergo Brownian motion.



To mimic the motor-driven filament sliding in actomyosin networks we describe the motors as generating anti-correlated kicks on pairs of crosslinks along their lines of centers. Assuming a fixed kick step size $l$, consistent with the nearly periodic structure of the actin filaments, an anti-correlated kick pair acting on nodes $i$ and $j$ can be represented by $(\vec l_{ij},\vec l_{j\,i})=l(\hat r_{ij},-\hat r_{ij})$, where $\hat r_{ij}$ is a unit vector pointing from node $i$ to node $j$.
These anti-correlated kick pairs with equal size automatically satisfy momentum conservation on the macroscopic scale. Yet if we include explicitly the aqueous environment in which the actomyosin network is immersed, hydrodynamic interactions between the nodes via the solvent should be taken into account. These interactions might modify the current simplified picture and counteract any motor-induced force imbalance on individual nodes, thus validating momentum conservation on the microscopic scale.

Dynamical evolution of the many-particle configuration $\ris$ due to these motor-driven nonequilibrium processes can be described by a master equation $\partial\Psi/\partial t=\Lne\Psi$ for the configurational probability density $\Psi(\ris;t)$ with
$\hat{L}_\textrm{NE}\Psi(\{\vec r\},t)=\int\Pi_i d\vec{r'_i}[K(\{\vec{r'}\}\rightarrow\{\vec{r}\})\Psi(\{\vec{r'}\},t)-K(\{\vec{r}\}\rightarrow\{\vec{r'}\})\Psi(\{\vec{r}\},t)]
$
where the integral kernel $K(\rips\rightarrow\ris)$ encodes the probability of transitions between different crosslink/node configurations. Our earlier description \cite{Teff ppr, spontaneous motion} of the motor kicking rate, $k$, still applies to the current case of correlated kicks, i.e.,
\begin{equation}\label{kinetic rate}
k=\kappa[\Theta(\Delta U)\exp(-s_u\beta\Delta U)+\Theta(-\Delta U)\exp(-s_d\beta\Delta U)],
\end{equation}
where $\kappa$ is the basal kicking rate and $s_u$($s_d$) denotes motor susceptibility to energetically uphill (downhill) moves, except that the free energy change $\Delta U$ now arises from \textit{pairs} of displacements.
Explicitly we write
\begin{eqnarray}
\nonumber&&\Lne\Psi(\ris;t)=\frac{1}{2}\kappa\sum_i\sum_j C_{ij}\int d\rip\int d\rjp\\ \nonumber
&&\times \Big\{\delta(\ri-\rip-\vec l_{ij})\delta(\rj-\rjp+\vec l_{ij})w\left[U(\cdots,\rip,\cdots,\rjp,\cdots)-U(\cdots,\ri,\cdots,\rj,\cdots)\right]\Psi(\rips;t)\\ \nonumber
&&-\delta(\ri-\rip+\vec l_{ij})\delta(\rj-\rjp-\vec l_{ij})w\left[U(\cdots,\ri,\cdots,\rj,\cdots)-U(\cdots,\rip,\cdots,\rjp,\cdots)\right]\Psi(\ris;t)
\Big\}.
\end{eqnarray}
The factor $1/2$ avoids double counting in the summation over all pairs. The quantity $C_{ij}$, much like an element of a contact map in description of protein structures, defines whether the node pair ($i,j$) is connected by an active bond and thus subject to anti-correlated displacements ($\vec l_{ij}, -\vec l_{ij}$): $C_{ij}=C_{ji}=1$ for motor-bonded pairs while $C_{ij}=C_{ji}=0$ for non-bonded pairs. Our description of the rates gives
$w[U_i-U_f]=\Theta(U_f-U_i)\exp[-s_u\beta(U_f-U_i)]+\Theta(U_i-U_f)\exp[-s_d\beta(U_f-U_i)]$.

Assuming symmetric motor susceptibility, i.e. $s_u=s_d=s$, one finds more simply
\begin{eqnarray}\label{Lne}
\nonumber&&\Lne\Psi(\ris;t)=\frac{1}{2}\kappa\sum_i\sum_j C_{ij}\\ \nonumber
&&\times \Big\{e^{-s\beta\left[U(\ri,\rj)-U(\ri-\vec l_{ij},\rj+\vec l_{ij})\right]}
\Psi(\{\cdots,\rip=\ri-\vec l_{ij},\cdots,\rjp=\rj+\vec l_{ij},\cdots\};t) \\
&& -e^{-s\beta\left[U(\ri+\vec l_{ij},\rj-\vec l_{ij})-U(\ri,\rj)\right]}\Psi(\{\cdots,\ri,\cdots,\rj,\cdots\};t)\Big\}.
\end{eqnarray}
We assume that kicks on different node pairs at any time are uncorrelated.
The rates of possible kicking events depend on the \textit{instantaneous} node configuration reflecting an assumed Markovian character of the dynamics. There is no angular average due to the definiteness of kicking directions for a given configuration.

Note that the motor power strokes and thus kick steps are discrete occurring in a stochastic fashion. The correlated motions pull in slack locally while pulling taut neighboring filaments until a global balance is reached or a macroscopic collapse occurs, depending on whether the motors are downhill-prone (with a large positive $s$) or load-resisting (with a small or negative $s$), respectively.
The latter may be relevant to the contractile rachet-like behavior \cite{contractile ratchet} that operates to \emph{incrementally} drive cell shape change and deform tissues.

\subsection{Numerical translation: dynamic Monte Carlo simulation}


To realize the finite-jump Markov process described by the (chemical) master equation (Eq.~\ref{Lne}) we performed dynamic Monte Carlo \cite{MC} simulations on the model actomyosin network (Fig.~\ref{schematic}).
In these simulations we generated initially a three-dimensional random lattice of volumeless nodes
(mimicking the crosslinking proteins) and connected the nearest-neighbor nodes (defined by the first shell of
the pair distribution function) with nonlinear elastic bonds \cite{JCP} (mimicking the actin filaments)
at a given probability $P_c$. We then distributed the myosin motors uniformly to the bonds at a given probability $P_a$ and obtained an active bond map. Considering the anti-correlated kicks along individual active bonds as chemical reaction channels, we adopted a stochastic simulation algorithm \cite{MC} to execute the moves following the stochastic process defined by the model motor kicking noise (Eq.~\ref{kinetic rate}). For a sufficiently large system, kicking events on different node pairs are effectively decoupled (consistent with summing over independent reaction channels in the master equation). Intermediate thermal moves between successive chemical moves obey Brownian dynamics \cite{Brownian dynamics} implemented via the position Langevin equation.


The bond properties are given by the elasticity onset or relaxed length of actin filaments $L_e=1.2$ and the effective stretch modulus $\beta\gamma=2$. Since the relaxed length is larger than the mean node separation (set as the length unit), the initial homogeneous (but amorphous) network has a considerable fraction of floppy bonds. We assumed a relatively high basal kicking rate ($\kappa=0.01$) and a large kick step size ($l=0.2$) such that the dimensionless motor activity (defined as $\kappa l^2/D_0$ where $D_0$ is the thermal diffusion constant) is close to $1$ and thus the strength of chemical noise is at least comparable to that of the thermal noise. The system size is $N=256$ and periodic boundary conditions are applied.
The relevant biophysical parameters to vary include network connectivity ($P_c$) as well as motor concentration ($P_a$) and susceptibility ($s$).



\section{Main Findings}

\subsection{Role of motor susceptibility ($s$) and concentration ($P_a$) in contractile behavior}

\begin{figure}[htb]
\centerline{\includegraphics[angle=0, scale=0.75]{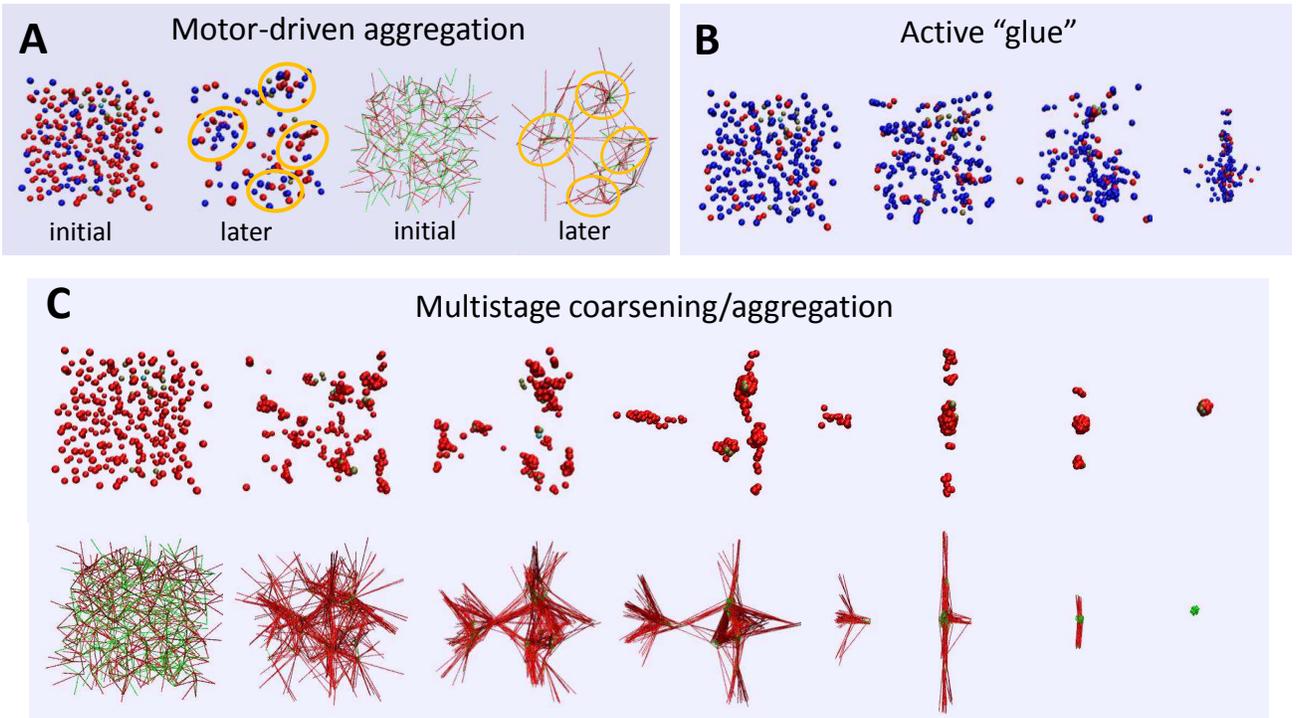}}
\caption{Illustration of how motor concentration $P_a$ and susceptibility $s$ contribute to their collective behavior. (a) Left: motor-driven aggregation of active nodes (red spheres) among the passive nodes (blue spheres); right: formation of dense floppy clumps (concentrated short green lines) inter-connected by tense bonds (long red lines). Circles mark the corresponding regions of node aggregation and bond collapse. $P_c=0.2, P_a=0.5, s=1$. (b) Active nodes tend to ``glue" together passive nodes and their aggregates. $P_c=0.5, P_a=0.2, s=0$. (c) Multistage coarsening/aggregation of active condensates driven by high-concentration ($P_a=1$) uphill-prone ($s=-0.5$) motors. $P_c=0.5$. Shown are the temporal evolution of the node configuration (upper row) and of the corresponding network structure (lower row). }
\label{Illustration_Pa_s}
\end{figure}

We first study how the concentration ($P_a$) and susceptibility ($s$) of motors contribute to the collective behavior. As illustrated in Fig.~\ref{Illustration_Pa_s}, depending on the specific combinations of $P_a$ and $s$, distinct nonequilibrium structures emerge. For a force-percolating network (i.e.\, one with connectivity beyond the percolation threshold) when partially motorized ($P_a<1$) under susceptible (large $s$) anti-correlated kicks, the active nodes (those with motor-attached bonds; shown as red spheres) begin to aggregate and tend to separate from the passive nodes (those with no active bonds; shown as blue spheres). The left panel of Fig.~\ref{Illustration_Pa_s}a shows snapshots of both the initial and later node configurations. The corresponding developed network structure (Fig.~\ref{Illustration_Pa_s}a right) exhibits clumps of floppy bonds (concentrated short green lines) connected by tense bonds (long red lines). The overall rigidity (i.e. homogeneity on large scales) of the structure is protected by susceptible motors which tune the balance between local bond contraction and neighboring bond stretching such that energetically unfavorable tense states are avoided.
At a low concentration of adamant motors (small $s$), however, active nodes and their aggregates tend to ``glue" together progressively the passive nodes and their condensates (Fig.~\ref{Illustration_Pa_s}b).
A finite spatial extent of the condensate and a non-vanishing fraction of taut bonds remain due to the insufficient cooperativity (low $P_a$) between local aggregation events.
More dramatically, significant spatial heterogeneity forms when the system is driven by a large number of uphill-prone motors with negative susceptibility: the cooperative action of load-resisting motors induces a multistage aggregation and coarsening of the nodes (Fig.~\ref{Illustration_Pa_s}c upper row) finally leading to a macroscopic contraction of an initially homogeneous network into a dense clump of buckled filaments (Fig.~\ref{Illustration_Pa_s}c lower row).
This multistage coarsening process involves three steps:
(1) local bond contraction and node aggregation giving floppy clumps connected by tense filaments; (2) coarsening of the aggregates leading to filament alignment and formation of tense bundles; (3) coalescence of the larger aggregates into a single condensate accompanies collapse of the tense bundles into a floppy clump. The aligned tense bundles formed before the eventual collapse constitute a taut state that can generate contractile forces.

We will investigate the scenario of motor-driven aggregation and pursue its analogy to arrested phase separation later. In the present work we will focus on the regime for macroscopic contraction.

To identify the required motor properties for active contractility, we performed many simulations to obtain the evolution (in Monte Carlo time $t_\textrm{MC}$) of the statistical characteristics for a series of motor susceptibilities (Fig.~S1a) and concentrations (Fig.~S1b).
These measures consistently indicate the existence of a threshold motor concentration ($(P_a)_\textrm{th}$) and a threshold susceptibility ($s_\textrm{th}$) for the onset of macroscopic contraction within the simulation time window.
When $s$ is less than $s_\textrm{th}$ at an intermediate $P_a$ value (Fig.~S1a: $s\leq0.02, P_a=0.5$) or for $P_a>(P_a)_\textrm{th}$ at a small $s$ value (Fig.~S1b: $P_a\geq0.3, s=0$), the fraction of taut bonds drops to essentially zero, indicating that an initially homogeneous percolating network collapses into a floppy clump.
More interestingly, the total energy first rapidly rises and reaches a maximum before falling to zero with the fraction of taut bonds. This nonmonotonic behavior suggests that the system first works against an energy barrier due to the formation of a \emph{transient} tense state (having highly stretched bundles induced by adamant or uphill-prone motor kicks), and then cooperative action of sufficiently load-resisting motors drives the system over the barrier to allow energetically downhill moves via subsequent coarsening (as shown in Fig.~\ref{Illustration_Pa_s}c). The rapid increase and saturation of the mean squared displacement (MSD) mirrors the evolution of the fraction of taut bonds and results from the formation of a single isolated floppy clump. Larger $s$ and/or higher $P_a$ yields a lower barrier to the collapsed state (as indicated by arrows in Fig.~S1). Moreover, lower $P_a$ necessitates having a smaller $s$ to induce macroscopic contraction. In other words, weaker motor cooperativity requires a stronger load-resisting tendency to trigger contractile instability.
When $P_a<(P_a)_\textrm{th}$ and/or $s>s_\textrm{th}$ the structure remains homogeneous except for modest \emph{local} node aggregation and network deformation.

\subsection{State diagram for active contractility: Interplay of network connectivity and motor cooperativity}

We now examine the interplay of network connectivity ($P_c$) and motor concentration ($P_a$) in forming contractile structures. The equivalence between the crosslink concentration and the fraction of bonded neighboring crosslinks $P_c$ (both are proportional to the number of inter-crosslink segments and define the number of bond constraints), and between the motor concentration and the fraction of active bonds $P_a$ (both are proportional to the number of crosslinks subject to motor kicks and determine the spatial cooperativity between motors) allows us to compare our state diagram constructed on the $P_a$-$P_c$ plane with the experimental result \cite{active contractility} shown for the parameter space of concentration ratios [myosin]/[actin] versus [$\alpha$-actinin]/[actin].

\begin{figure}[htb]
\centerline{\includegraphics[angle=0, scale=0.6]{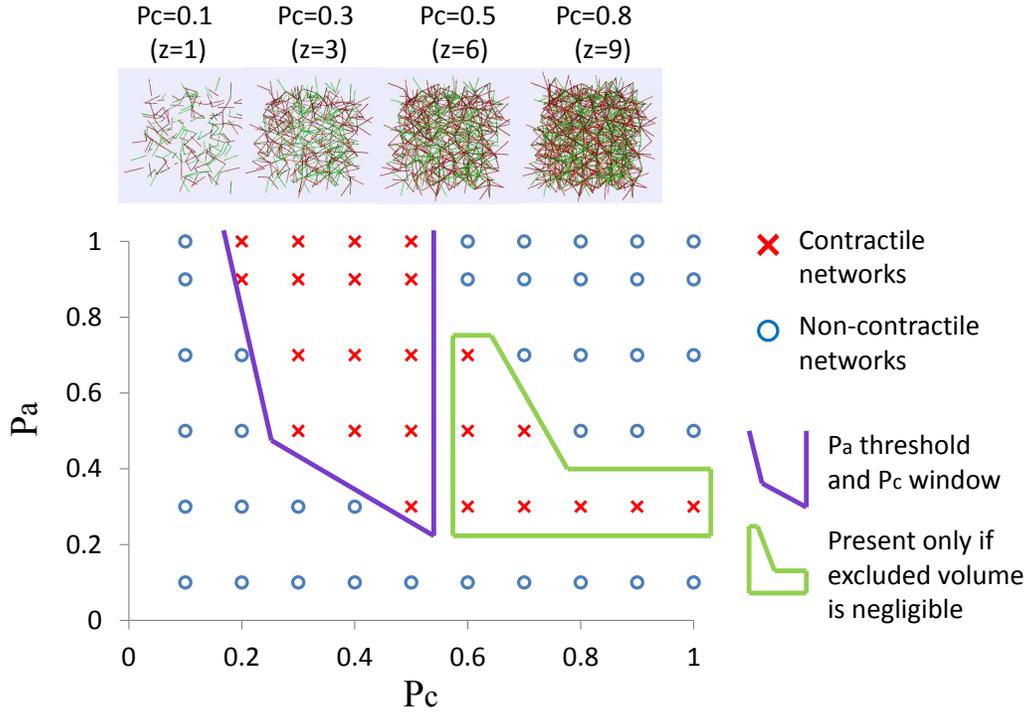}}
\caption{State diagram showing the dependence of active contractility on motor concentration $P_a$ and network connectivity $P_c$. $s=0.01$. Red crosses denote contractile networks and blue circles denote non-contractile networks. Top row displays the initial homogeneous network structures with increasing connectivity $P_c$ and average number of bonded neighbors $z$. Macroscopic contraction occurs in the two framed regions: (1) purple open frame: intermediate $P_c$ and above-threshold $P_a$; (2) green closed frame: high $P_c$ and intermediate $P_a$.}
\label{diagram_Pa_Pc}
\end{figure}

We present our state diagram showing the dependence of macroscopic contractility on the network connectivity and motor concentration at a small $s$ value ($s=0.01$) in Fig.~\ref{diagram_Pa_Pc}. Red crosses denote contractile networks while blue circles denote non-contractile networks. By ``contractile" we mean a complete collapse of an initially homogeneous network into a floppy clump within $10^7$ MC steps, monitored by the vanishing of total potential energy and the fraction of taut bonds.
On top of the figure, we display the initial network structures for several typical values of network connectivity $P_c$ with the average number of bonded neighbors $z$.
As observed experimentally,\cite{active contractility} we identify a threshold motor concentration and a window of network connectivity for active contractility, which define a parameter region as marked by the purple open frame.
The new feature is a small window of motor concentration for contraction at high connectivity (marked by a green closed frame).
These two aspects vividly demonstrate the interplay of network connectivity and motor concentration for global contraction:
at any connectivity beyond the percolation threshold ($P_c\geq0.2$), a sufficiently high motor concentration is required to achieve cooperativity among local contraction events; on the other hand, since the bond constraints are strong at high connectivity, motor concentration cannot be too high since force asymmetry (imbalanced tug-of-war) is necessary to trigger local contraction. In the \textit{in vitro} experiments,\cite{active contractility} no macroscopic contraction was observed on the hour time scale at high crosslink concentrations. This observation is not incompatible with our results since inclusion of excluded volume effects should give a dramatic slow-down of the contractile dynamics owing to jamming and/or glass transition which would account for the absence of observable contraction on laboratory time scales.

Outside the framed regions, at low motor concentration and/or low connectivity as well as at high motor concentration and high connectivity, there is no macroscopic contraction.
We illustrate the failure of contractility at low connectivity (Fig.~S2a) or low motor concentration (Fig.~S2b). For $P_c$ as low as $0.1$, the average number of bonded neighbors is no greater than $1$, lack of tension percolation thus prevents \emph{global} contraction. This becomes more obvious as we increase the motor concentration; an increasing trend of \emph{local} collapse with rising $P_a$ value is apparent (Fig.~S2a), leading to more compact aggregates and disconnected floppy clumps (several typical spots have been circled for the highest $P_a$ case).
At very low motor concentration $P_a=0.1$ (Fig.~S2b), formation of sparse and small active-node foci did not dramatically reshape the network, since rare and separate contraction events are insufficient to trigger global contractile instability.

A quantitative demonstration of the interplay between $P_c$ and $P_a$ is given in Fig.~S3. Macroscopic contraction occurs either for high motor concentration (upper row, $P_a=0.7$) at intermediate connectivity ($P_c=0.3, 0.5$) or for high connectivity (lower row, $P_c=0.7$) at intermediate motor concentrations ($P_a=0.3, 0.5$). Lack of percolation at $P_c=0.1$ is signalled by the diffusive behavior of the mean square node displacement MSD linearly increasing with $t_\textrm{MC}$ (red line indicated by arrow in Fig.~S3a right panel) since the disconnected aggregates merely undergo thermal motion. On the other hand, a small steady value for the MSD at $P_a=0.1$ (red line indicated by arrow in Fig.~S3b right panel) reflects the low cooperativity which causes no more than modest local distortions. A balanced tug-of-war at high $P_c$ and high $P_a$ disfavors even local deformations yielding a lower fraction of taut bonds, lower energy and smaller MSD as $P_a$ increases. This can be seen by comparing the measures for $P_a=0.1$ and $P_a=0.7$ at $P_c=0.7$ in Fig.~S3b.

\subsection{Contrast with the case for uncorrelated kicks}

Our earlier study of a model cytoskeleton with uncorrelated isotropic kicks acting on individual nodes \cite{spontaneous motion} revealed that force percolation and mechanochemical coupling due to susceptible motors can conspire to maintain a spontaneous flow, whereas adamant motor kicks promote fluidization (characterized by a vanishing localization strength of the nodes and formation of a disordered tense network). Under anti-correlated kicks acting along the lines of centers of motor-bonded node pairs, however, no vectorial flow transition is found even for a very large kick step size ($l\simeq0.5$). This is probably because the restrictiveness of (local) kicking directions for a given configuration impedes a global concerted \emph{net} movement of the whole lattice. Instead, balanced local contraction and neighboring stretch in the presence of force percolation results in a network of floppy spots connected by tense filaments which remains homogeneous on large scales and bears some resemblance to arrested phase separation occurring in low-packing-fraction physical gels with strong short-range attractions \cite{colloidal gels}.
At intermediate connectivity, force imbalance sensed by susceptible motors still induces phase separation into large floppy clumps connected by taut inter-clump bonds yet oscillations are no longer found.
Fluidization is replaced by global contraction at above-threshold motor concentration.

\begin{figure}[htb]
\centerline{\includegraphics[angle=0, scale=0.55]{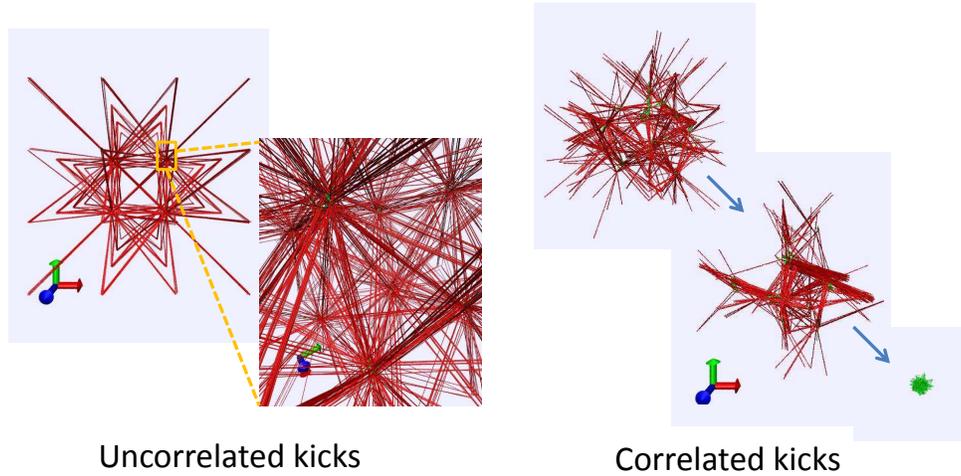}}
\caption{Aster versus condensate. $P_c=0.5, P_a=1, s=-0.5$. Left: Under uncorrelated kicks, an initially disordered and homogeneous network self-organizes into highly tense and ordered ``asters", composed of tense bundles radiating from the junctions where floppy bonds concentrate (see the zoom-in image). Right: Driven by anti-correlated kicks, an initially homogeneous force-percolating network first develops transient tense states consisting of highly-stretched bundles, but then abruptly collapses into a single floppy clump.}
\label{aster_vs_condensate}
\end{figure}

When the uncorrelated-kicking system was driven by uphill-prone motors with negative susceptibility, aster-like patterns formed (Fig.~\ref{aster_vs_condensate} left). Instead, now, multistage coarsening and eventual macroscopic collapse occur (Fig.~\ref{aster_vs_condensate} right), after surmounting a high energy barrier caused by transient tense states with stretched bundles. Comparing the bond tension patterns for these two cases in Fig.~\ref{aster_vs_condensate}, we see that global contraction requires \emph{correlated} movements that locally buckle a filament yet impose strain on the bonds at \emph{both} ends. Without this correlation the aster pattern cannot collapse.
We display in Fig.~S4 the snapshots in the course of aster formation which clearly demonstrate that node aggregation and coarsening leads to progressively more tight filament bundling and bundle alignment, a mechanism that operates for both the correlated- and uncorrelated-kick cases.

\section{Conclusion and Discussion}

We have simulated a microscopic model for the actomyosin cytoskeleton as a motorized cat's cradle that combines the asymmetric load response of individual actin filaments with correlated motor-driven events. This model reproduces the dependence of active contractility on microscopic parameters observed in reconstituted actomyosin networks. The simulations allow us to identify several necessary conditions for active contractility: highly asymmetric load responses of the filaments are needed for local contractile behavior under anti-correlated kicks, and a minimal network structure beyond percolation threshold is required to propagate local contraction. Sufficiently cooperative load-resisting motors manage to drive the system through energy-costly intermediate states and incorporate local buckling events into macroscopic contraction via a multistage coarsening process.

This numerical study provides an explanation for the formation and contractile dynamics of disordered condensed state of actomyosin \textit{in vivo}.  The fact that such a simplified model is able to mimic cellular self-organization states and their contractile dynamics suggests that purely physical interactions contribute to the regulation of cell and tissue morphogenesis. Nevertheless specific biochemical signaling events \cite{Rho regulation} certainly contribute to the \emph{localized} assembly and activation of myosin foci such as occurs in cleavage furrow during cytokinesis and in wound borders.


It is clearly necessary in the future to take into account the excluded volume effect which is expected to dramatically slow the contractile dynamics at high crosslink concentrations, thus accounting for the failure to see contractility on laboratory time scales.
Long-range hydrodynamic interactions between the node pairs may also change the qualitative physics.
Hydrodynamic correlations should facilitate propagation of nearby kicking events and break the continuous symmetry so as to allow spontaneous directed motion with motor-driven hydrodynamics-mediated pulsed contractions.
Finally, in vivo one must incorporate motor attachment and detachment and crosslink binding and unbinding for complete realism.




Support from the Center for Theoretical Biological Physics sponsored by the National Science Foundation (Grant PHY-0822283) is gratefully acknowledged.

\newpage

\newpage

\centerline{\includegraphics[angle=0, scale=0.6]{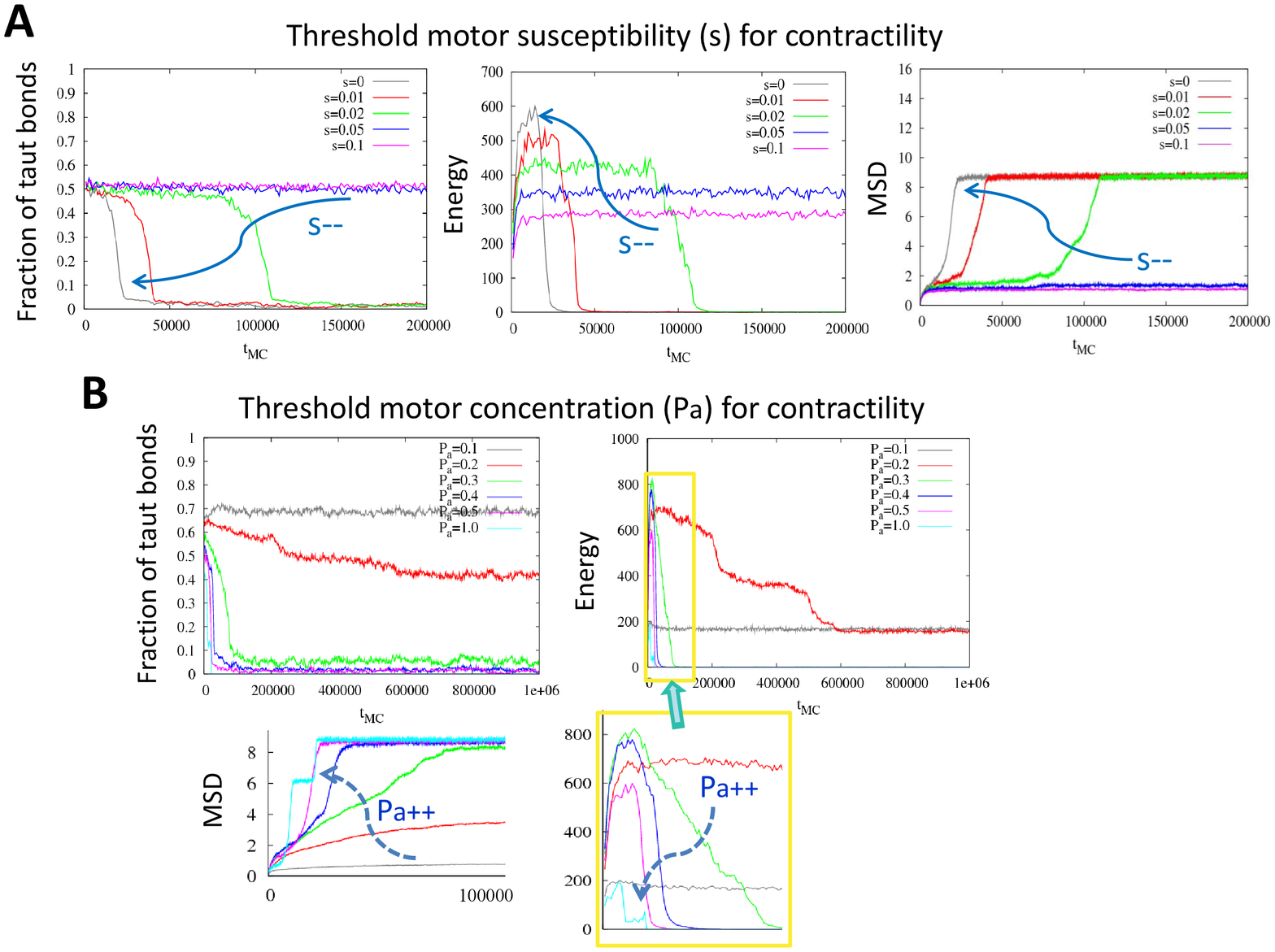}}

\noindent
{\bf Fig.~S1 $|$ Thresholds for active contractility.} (a) Evolution of the statistical measures for a series of motor susceptibilities ($s=0, 0.01, 0.02, 0.05$ and $0.1$). $P_c=0.5, P_a=0.5$. When $s\leq0.02$ the fraction of taut bonds and the total energy drop to essentially zero after surmounting an energy barrier due to tense intermediates. Mean square displacement (MSD) mounts to a plateau as a consequence of the formation of a single floppy clump. Smaller $s$ yields a higher barrier and faster collapse (indicated by arrows). (b) Evolution of the measures for a series of motor concentrations ($P_a=0.1, 0.2, 0.3, 0.4, 0.5$ and $1$). $P_c=0.5, s=0$. When $P_a\geq0.3$ global contraction occurs. Larger $P_a$ leads to lower barrier and faster collapse (indicated by arrows).

\newpage

\centerline{\includegraphics[angle=0, scale=0.6]{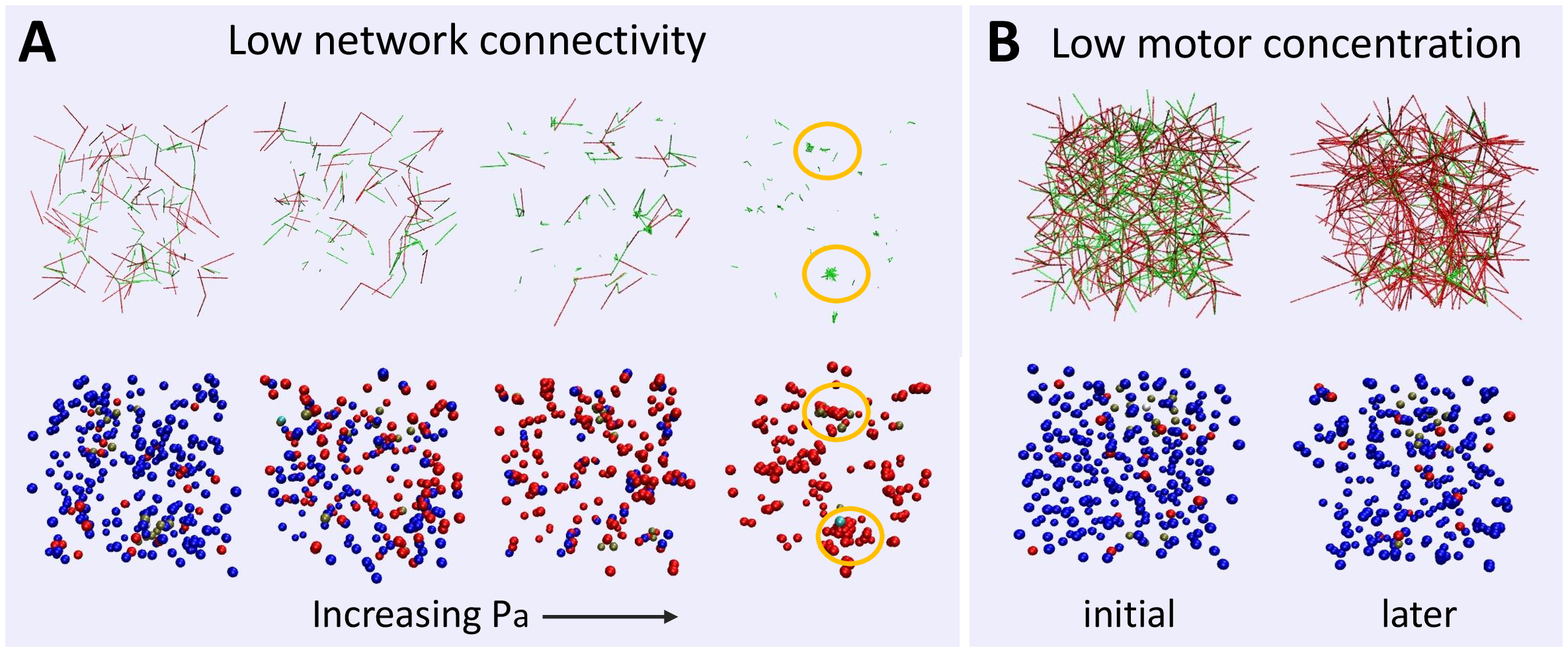}}

\noindent
{\bf Fig.~S2 $|$ Illustrations of the failure of contractility at low network connectivity or low motor concentration.}  $s=0$. (a) $P_c=0.1$. Lack of tension percolation due to low degree of bonding prevents global contraction. As $P_a$ increases (left to right: $P_a=0.1, 0.5, 0.8$ and $1$) system exhibits increasing trend of local aggregation (lower row) resulting in an increased fraction of floppy bonds (upper row). When driven by high-concentration ($P_a=1$) adamant motors, the initially homogeneous network exhibits local collapses into disconnected clusters of buckled filaments (circles mark the typical regions). (b) $P_a=0.1, P_c=0.5$. Sparse and modest local network distortion (upper) and node aggregation (lower) are insufficient to trigger global contractile instability.

\newpage

\centerline{\includegraphics[angle=0, scale=0.6]{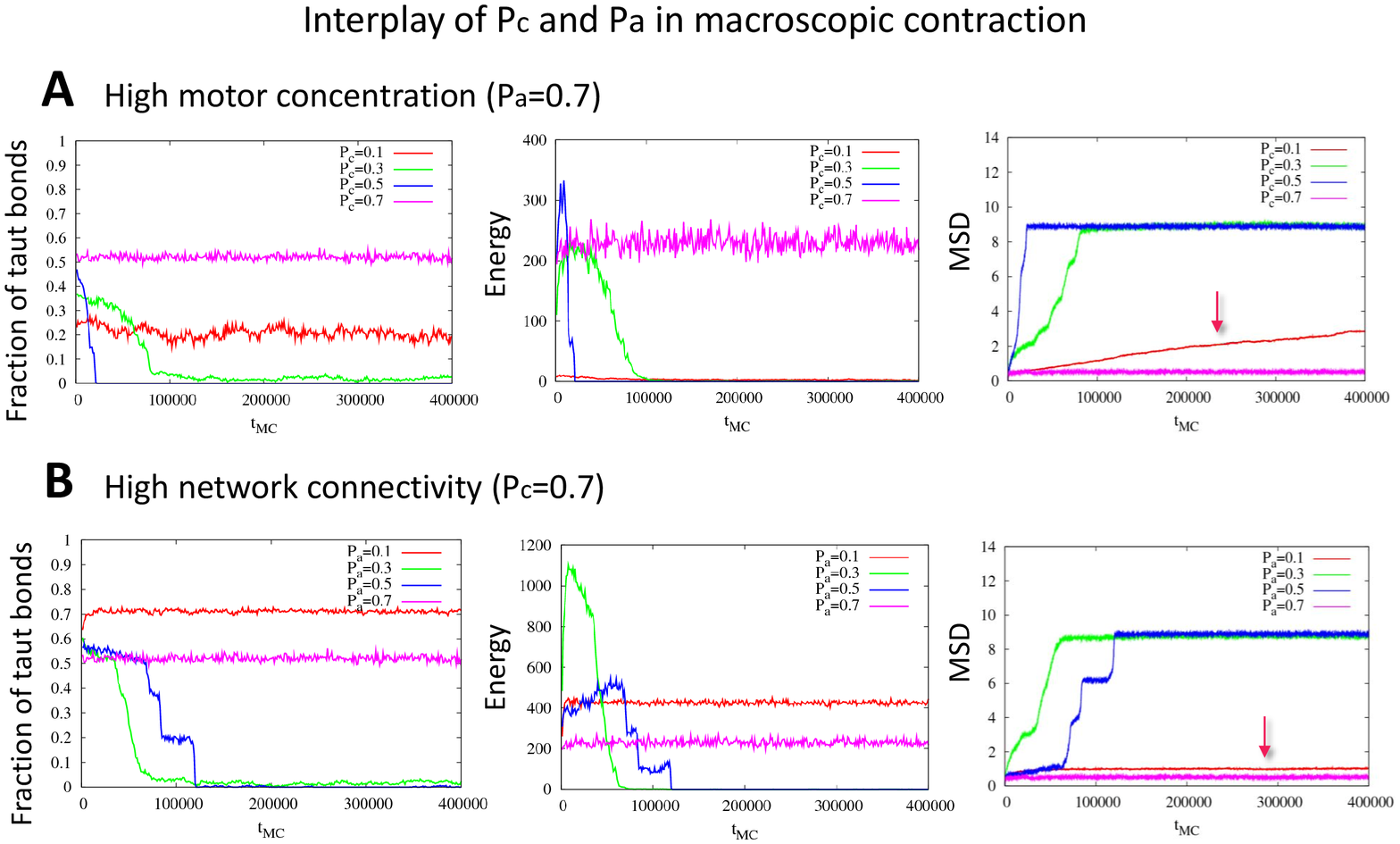}}

\noindent
{\bf Fig.~S3 $|$ Interplay of network connectivity and motor concentration in macroscopic contraction.}
$s=0.01$. (a) $P_a=0.7$, $P_c=0.1, 0.3, 0.5, 0.7$: contraction occurs for intermediate $P_c$. Larger $P_c$ yields higher ``barrier" to collapsed state since more bond constraints results in more intense tug-of-war. (b) $P_c=0.7$, $P_a=0.1, 0.3, 0.5, 0.7$: contraction occurs for intermediate $P_a$. Larger $P_a$ lowers the barrier since cooperativity between local contraction events is enhanced.

\newpage

\centerline{\includegraphics[angle=0, scale=0.6]{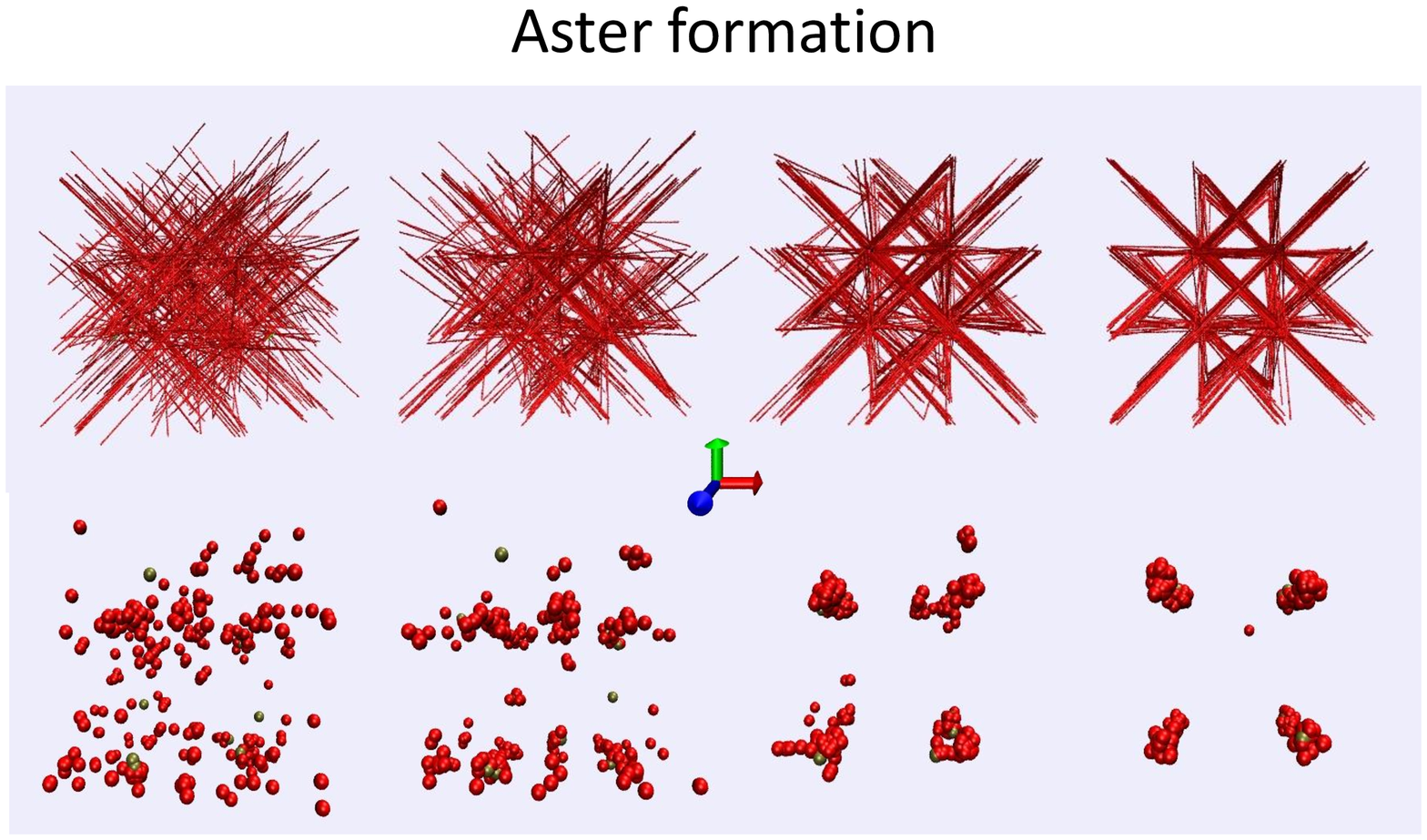}}

\noindent
{\bf Fig.~S4 $|$ Snapshots in the course of aster formation.}
Upper row shows the bond structure: filament bundling evolves from loose to tight. Lower row presents the corresponding node configuration: aggregation toward the corners of the simulation box becomes progressively more compact. $P_c=0.3, P_a=1, s=-0.5$.

\end{document}